\documentclass[aps,twocolumn,showpacs,preprintnumbers,amsmath]{revtex4}
\usepackage{dcolumn}
\usepackage{bm}
\usepackage{graphicx}
\usepackage{bm}
\begin{document}
\title{Variation of T$_g$ with magnetic field in a magnetic glass}
\author{Archana Lakhani, Pallavi Kushwaha, R. Rawat, Kranti Kumar, A. Banerjee and P. Chaddah.}
\affiliation{UGC-DAE Consortium for Scientific Research (CSR)\\University Campus, Khandwa Road\\
Indore-452001, India.}
\date{\today}

\begin{abstract}
Glass-like arrest has recently been reported in various magnetic materials. As in structural glasses, the kinetics of a first-order transformation is arrested while retaining the higher-entropy phase as a non-ergodic state. We show visual mesoscopic evidence of the irreversible devitrification of the arrested antiferromagnetic-insulating phase in $Pr_{0.5}Ca_{0.5}Mn_{0.975}Al_{0.025}O_3$ to its equilibrium ferromagnetic-metallic phase with isothermal increase of magnetic field, similar to its iso-field devitrification on warming. The slope of $T_g$ vs magnetic field in half-doped manganites is shown to have a sign governed by Le Chatelier's Principle. 
\end{abstract}


\maketitle
Glasses form when kinetics is arrested below a temperature $T_g$, preserving the high-temperature structure while avoiding the first-order liquid-solid transformation. The critical cooling rate required for vitrification has been shown to fall sharply as the ratio of $T_g$ to the melting temperature $T_m$ rises \cite{Greer, Inou}; the suppression of $T_m$ with increasing pressure has been used to vitrify monoatomic germanium at achievable cooling rates \cite{Bhat}. 

Glass like arrest of kinetics across first-order magnetic transition has been reported in materials ranging across intermetallics and colossal magnetoresistance manganites \cite{Chat, Bane, Wu, Macia, Roy, Chad, Rawat}, while preserving the high-temperature phase as a non-ergodic state. These are termed as `magnetic glass' where a first-order magnetic transition is inhibited by lack of kinetics, and one obtains glass-like arrested states (GLAS). The kinetic arrest in these GLAS is reflected in the decay of the arrested state following Kohlrauch-Williams-Watts form as one approaches the glass transition temperature $T_g$ \cite{Chat, Roy}, and in the arrested state showing devitrification on warming \cite{Bane, Roy, Chad}. Further, these magnetic glasses have an excess specific heat that varies linearly with T as in conventional glasses including orientational glasses \cite{spht}. Depending on the system, the GLAS can be antiferromagnetic \cite{Bane, Wu, Macia, Roy} or ferromagnetic \cite{Chat, Bane, Chad, Rawat}. The half doped collosal magnetoresistance (CMR) manganites have an important advantage over other materials because the conductivity changes drastically along with magnetic order across the transition. While a decrease in global magnetization of the sample can be interpreted as either reduction of moment in ferromagnetic metallic (FM-M) phase, or as part transformation of FM-M to antiferromagnetic insulator (AF-I), a simultaneous measurement of conductivity provided a clear choice between the two alternatives because of the orders of magnitude resistivity changes associated with the metal (M) to insulator (I) transition in the latter case. A visual demonstration of both these types of changes i.e. change in the phase fraction  \cite{Wu} as well as change in the global magnetization can be captured by mesoscopic measurements with a magnetic force microscope (MFM).      

It has been shown in recent literature \cite{Bane, Macia, SharP} that while an equilibrium phase coexistence is due to quenched impurities that lead to the spread of the local transition temperatures (where local means over length scales of the order of the correlation length), and a consequent rounding of the first order phase transition (FOPT), the static phase coexistence persisting to lower temperature without change in phase fraction is due to glass-like arrest of the transformation kinetics (recent work \cite{SharP} identifies a possible kinetics that is frozen at $T_g$). $Pr_{0.5}Ca_{0.5}Mn_{0.975}Al_{0.025}O_{3}$ (PCMAO) is one of the extensively studied members of this class of magnetic glasses, whose GLAS is an AF-I, while the low-temperature equilibrium state is a FM-M \cite{Bane, Bane1}. The fraction of the glass-like AF-I phase at a given H increases by making the cooling field $H_C$ smaller \cite{foot}. The AF-I phase obtained as a homogeneous state on cooling PCMAO in H=0, shows all the characteristics of a glassy state including devitrification. A specially designed `cooling and heating in unequal fields' (CHUF) protocol has been used to show that devitrification occurs whenever the sample is warmed in a magnetic field ($H_w$) that is higher than the field ($H_c$) it was cooled in ; a reentrant AF-I to FM-M to AF-I transition is seen on heating as devitrification at $T_g$ is followed by the first order transition at $T^{**}$ analogous to melting at $T_m$ \cite{Bane1}. Here we show that while $T^{**}$ rises with increasing H (as expected), $T_g$ falls with rising H. The latter is a new result in that the variation of $T_g$ with the second variable (usually pressure) is not normally investigated. \textit{The dependence is consistent with the applicability of Le Chatelier's Principle beyond its stated domain in that devitrification is a non-equilibrium process.} This conclusion is supported by mesoscopic MFM measurements showing that devitrification takes place with the isothermal increase of H, similar to devitrification observed on heating. 

In this study we have used the same PCMAO sample as in previous studies \cite{Bane, Bane1, Nair}. The preparation and characterization details for this sample can be found in \cite{Nair}. The magnetization measurements were performed using a 14 Tesla physical property measurement system-vibrating sample magnetometer (PPMS-VSM), M/s. Quantum Design, USA. For MFM measurements sample was polished to mirror finished surface by repeated grinding and polishing. Magnetic imaging was carried out using  Low temperature-High Field Magnetic Force Microscope from M/s. Nano-Magnetics Instruments, UK along with 9-Tesla superconducting magnet system from M/s. American Magnetics, USA. $NANOSENSORS^{TM} PPP-MFMR$ cantilevers with resonance frequency $\approx$70 kHz are used in the present study. The microscope utilizes the noncontact mode for magnetic as well as topographic scan. In this mode cantilever is tuned to its resonance frequency when it is in free condition or well away from the surface. Cantilever is brought closer to sample surface while monitoring the change in its resonance amplitude and frequency. During forward scan oscillation amplitude (i.e. root mean square of amplitude $V_{rms}$) is maintained constant by varying tip height thus giving topographic information. Reverse scan is used for magnetic imaging where cantilever is lifted by 50nm (called lift off) and follows the topography measured during forward scan. The change in amplitude ($V_{rms}$) during this scan gives the magnetic profile of the surface. 

In figure 1 we show measurements of M vs T following the CHUF protocol.
\begin{figure}[htbp]
	\centering
		\includegraphics{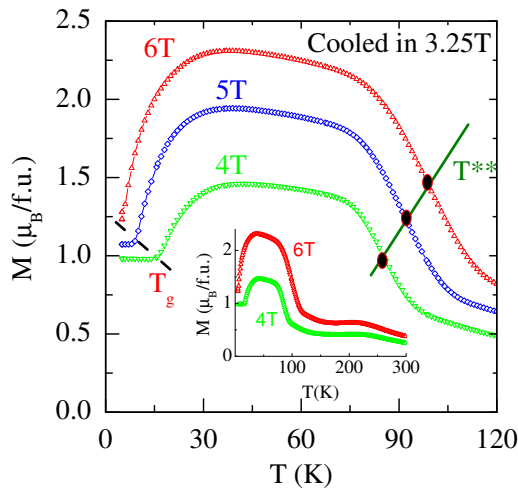}
	\caption{(color online) Magnetization (M) as function of temperature during warming in the presence of labeled magnetic field. For each of these measurements sample is cooled in the presence of 3.25 Tesla to 5 K and magnetic field changed to labeled field isothermally. Devitrification at low temperature and reverse transformation to AFM state is highlighted. Dashed line shows $T_g$ and $T^{**}$ line for these curves. See text for details. The inset shows the magnetization while warming in 4 and 6T after cooling in 3.25T for the complete temperature range.}
	\label{fig:Fig1}
\end{figure}
Cooling the sample from 300K to 5K in a cooling field $H_C = 3.25 Tesla$ results in coexisting phases with the AF-I phase fraction being about 70\%.
We now study the devitrification behavior of this magnetic glass as it is heated in different fields, in an attempt to see how T$_g$ depends on H. The sample is warmed in a field $H_w = 4 Tesla$ and the devitrification to FM-M phase is seen through a sharp rise in M that starts at 16 K and terminates at $\approx 40 K$, with the mid point ($T_g$ at 4 Tesla) being at $\approx 20K$. The re-entrant transition, corresponding to melting, is seen by a sharp fall in M, with the mid-point of the transition ($T^{**}$ at 4 Tesla) being at 84K. We again cool the sample from 300K to 5K in a cooling field $H_C = 3.25$ Tesla to obtain the same initial state with AF-I phase fraction being about 70\% \cite{Bane} and now warm from 5K after increasing the field to $H_w = 5$ Tesla. The devitrification for FM-M phase is again seen through a sharp rise in M that now starts at 9K, and terminates at about 35K, with the mid point ($T_g$ at 5 Tesla) being about 15K. The mid point of the re-entrant transition, corresponding to melting, ($T^{**}$ at 5 Tesla) is 90K. We again cool the sample from 300K to 5K in a cooling field $H_C = 3.25 Tesla$ and now warm from 5K after increasing the field to $H_w = 6$ Tesla. The devitrification to FM-M phase is again seen through a sharp rise in M that now starts at 5K itself and terminates at $\approx 30K$ with the mid-point ($T_g$ at 6 Tesla) being at $\approx 10K$. The mid point of the re-entrant transition, corresponding to melting, ($T^{**}$ at 6 Tesla) is 98K. We see that $T^{**}$ rises with increasing field so that an isothermal increases of H would, in a specific range of T, convert the AF-I state to FM-M. This is consistent with the qualitative condition imposed by Le Chatelier's Principle on a system in equilibrium; the value of $dT^{**}/dH$ should of course be consistent with the Clausisus Clapeyron relation since this is a first order transition. We now note that $T_g$ at 4 Tesla is 20K, while the mid point of $T_g$ at 6 Tesla is 10K. $T_g$ thus falls with increasing field, and an isothermal increase of H would, in a certain range of T, convert the glass-like AF-I phase to equilibrium FM-M phase by devitrification. This is consistent with the qualitative condition imposed by Le Chatelier's Principle in that increasing H takes the system to a state with higher M. Except that the glass-like AF-I phase is not a system in equilibrium; and devitrification is not an equilibrium process.

Devitrification under isothermal increase of H is not just a conjecture, and we present MFM data to support this.
In figure 2 we show MFM pictures of the re-entrant AF-I to FM-M to AF-I transformation with increasing T.
The sample is cooled from 300K to 5K in H=0, and the field is raised to 7 Tesla. We observe coexistence of antiferromagnetic (light color) and ferromagnetic (dark color) regions, at the micron level, as has been inferred under the ZFC condition \cite{Bane}. We now raise T and note an increase in FM fraction at T = 40K as shown in histogram in the bottom panel. Since the response of cantilever changes with temperature we have shown histograms of normalized $V_{rms}$ ($V_{norm}$). The re-entrant melting to antiferromagnetic phase is seen starting at 90K, and we have an almost homogeneous AF-I phase at 150K. This devitrification followed by `melting' with warming is as expected for any structural glass.
\begin{figure}[htbp]
	\begin{center}
	\includegraphics [width=8.5 cm]{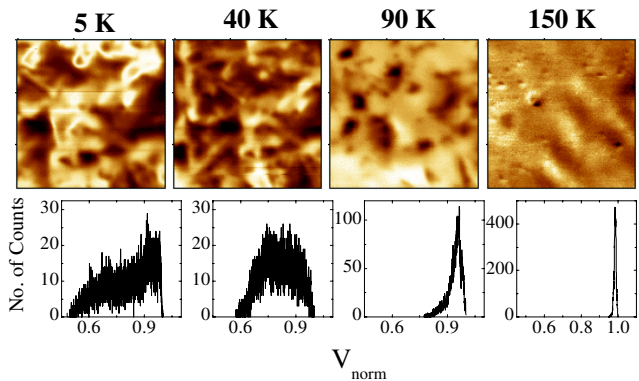}
	\end{center}
	\caption{(color online) MFM images of PCMAO in H = 7 Tesla taken during warming after sample is cooled to 5K in zero field. The magnetic field is applied isothermally at 5K. Dark color correspond to ferromagnetic regions. The reentrant transformation from AF-I to FM-M state (devitrification to FM as T is raised from 5 K to 40 K) to AF-I state (`melting' of FM to AF as T is raised from 40 K to 90 K) is observed. The distribution of normalized amplitude variation $V_{norm}$ (bottom panel) indicates coexisting FM and AF regions, while at 150 K the sample is AF and homogeneous. Frame size of each MFM scan is $17 \mu m \times 17 \mu m$.}
	\label{Figure 2}
\end{figure}

We now show in figure 3 the isothermal behavior of the glass. The sample is cooled to 6K in zero field, and H is then varied isothermally. At H = 1 Tesla, the sample is mostly antiferromagnetic, and a major fraction of the ferromagnetic phase is seen at H = 6 Tesla. We now reduce H to 1 Tesla (bottom panel), and find that there is no reverse transformation during the field reduction. The magnetic profile of the sample at 1 Tesla is identical to that of 6 Tesla
\begin{figure}[htbp]
	\begin{center}
	\includegraphics [width=8.5 cm]{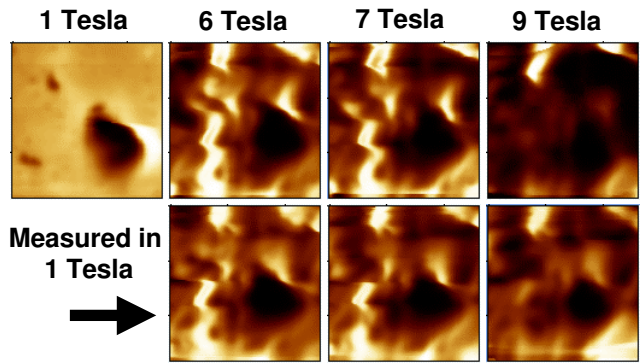}
	\end{center}
	\caption{(color online) MFM images of PCMAO with increasing magnetic field (top row) after sample is cooled to 6K in zero field. Dark patches correspond to ferromagnetic regions. Magnetic images were also taken at 1 Tesla (bottom row) after reducing magnetic field (isothermally at 6K) from that shown in top row. The magnetic profiles remain unchanged on reducing the magnetic field to 1 Tesla, showing irreversible isothermal transformation from AF-I to FM-M as expected in devitrification. Frame size of each MFM scan is $17 \mu m \times 17 \mu m$.}
	\label{Figure 3}
\end{figure}
image, which shows much higher ferromagnetic phase compared to first 1 Tesla image. Then H is raised to 7 Tesla with a visible increase in the ferromagnetic region; reducing H to 1 Tesla again causes no reverse transformation. We then raise H to 9 Tesla (our experimental limit) with a further visible increase in the ferromagnetic region; reducing H to 1 Tesla again causes no reverse transformation.

We thus show that increasing H isothermally at 6 K causes an irreversible conversion from antiferromagnetic to ferromagnetic phase, as expected in the process of devitrification. We have done magnetization measurement under the same protocol of cooling in zero field to 6 K and then varying H isothermally as above (figure 4).
\begin{figure}[htbp]
	\begin{center}
	\includegraphics{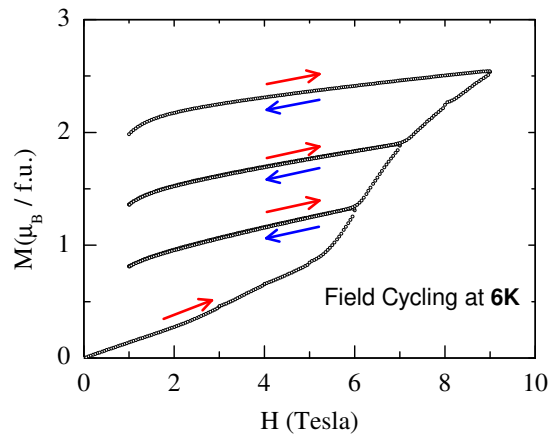}
	\end{center}
	\caption{(color online) Magnetization (M) vs. Magnetic field (H) at 6 K corresponding to MFM images shown in Figure 2. Irreversible isothermal transformation (devitrification) from AF-I to FM-M state is highlighted.}
	\label{Figure 4}
\end{figure}
In the initial field increasing cycle, the first order field induced AF-I to FM-M transition starts around 5.3 Tesla. After reaching the target field (H$_t$) of 6T the field is reduced to 1T but this 1T value is much higher than initial 1T value. It is important to note that the next field increasing cycle to 7T the magnetization follow the same path up to 6T. Again the 1T value after returning from H$_t$ of 7T gives higher values than the earlier. Similar feature is repeated for H$_t$ of 9T. Each return from progressively higher H$_t$ show higher value at 1T, there is no back conversion and field increasing path overlaps with the previous field decreasing path up to the previous H$_t$. These clearly indicate that field induced transformed FM-M phase fraction does not undergo reverse transformation with the reduction of field which is visually demonstrated in the maps of the bottom panel of Fig. 3. The minor decrease in magnetization between its value at H$_t$ and the return cycle is because of the global decrease in magnetization. This is also reflected in maps of Fig. 3, the shades of bottom panel are lighter than the top because of the global decrease in the magnetization without any change in the phase fractions. 
Thus the results shown in figure 4 are consistent with the MFM data of figure 3.

We have obtained data similar to that in Fig.1 following CHUF protocol for $Nd_{0.5}Sr_{0.5}MnO_3$ (NSMO),where the glassy phase is FM-M and it coexists with equilibrium AF-I phase \cite{Sdash}. We observe that $T_g$ rises and $T^{**}$ falls with increasing H in NSMO. This corresponds to isothermal devitrification of the non-equilibrium ferromagnetic phase to the equilibrium antiferromagnetic phase with decreasing H. This is again consistent with the qualitative condition imposed by Le Chatelier's Principle in that decreasing H takes the system to a state with lower M. Except that the glass-like FM-M phase is not a system in equilibrium; and devitrification is not an equilibrium process.

The ability to vitrify any liquid has been a challenging problem because of both the physics issues and also the potential for applications. Attempts to form glasses follow the empirical rule of increasing the ratio of $T_g/T_m$ and our result indicate how this ratio varies with the second variable that in our case is magnetic field. We have shown that the AF magnetic phase devitrifies with increasing H, and the FM magnetic glass devitrifies with decreasing H. This behavior is consistent with the magnetization of the magnetic glass being smaller or larger than that of the equilibrium or `crystal' state. Replace magnetization by density, and magnetic field by pressure, and this work on magnetic glasses may have relevance to structural glasses.

Department of Science \& Technology (DST), Government of India is acknowledged for funding the 14-Tesla PPMS-VSM at CSR, Indore.


\begin{thebibliography}{}
\bibitem[14] {foot}The fraction of the glass-like phase can of course be controlled, in a limited range, by varying the cooling rate. But here the fraction is controlled while keeping the cooling rate fixed.
\end{thebibliography}
\end{document}